\documentclass[twocolumn]{aastex63}

\received{January 31, 2021}
\revised{February 28, 2021}
\accepted{March 16, 2021}

\submitjournal{ApJL}

\shorttitle{Variability of Late-time Radio Emission in SLSN PTF10hgi}
\shortauthors{Hatsukade et al.}

\begin{document}

\title{Variability of Late-time Radio Emission in the Superluminous Supernova PTF10hgi}

\author[0000-0001-6469-8725]{B.~Hatsukade}
\affiliation{Institute of Astronomy, Graduate School of Science, The University of Tokyo, 2-21-1 Osawa, Mitaka, Tokyo 181-0015, Japan}
\email{hatsukade@ioa.s.u-tokyo.ac.jp}

\author[0000-0001-8537-3153]{N.~Tominaga}
\affiliation{Department of Physics, Faculty of Science and Engineering, Konan University, 8-9-1 Okamoto, Kobe, Hyogo 658-8501, Japan}
\affiliation{Kavli Institute for the Physics and Mathematics of the Universe (WPI), The University of Tokyo, 5-1-5 Kashiwanoha, Kashiwa, Chiba 277-8583, Japan}

\author[0000-0001-7449-4814]{T.~Morokuma}
\affiliation{Institute of Astronomy, Graduate School of Science, The University of Tokyo, 2-21-1 Osawa, Mitaka, Tokyo 181-0015, Japan}

\author[0000-0003-3932-0952]{K.~Morokuma-Matsui}
\affiliation{Institute of Astronomy, Graduate School of Science, The University of Tokyo, 2-21-1 Osawa, Mitaka, Tokyo 181-0015, Japan}

\author[0000-0003-4807-8117]{Y.~Tamura}
\affiliation{Department of Physics, Nagoya University, Furo-cho, Chikusa-ku, Nagoya 464-8602, Japan}

\author[0000-0002-8169-3579]{K.~Niinuma}
\affiliation{Graduate School of Science and Engineering, Yamaguchi University, Yoshida 1677-1, Yamaguchi, Yamaguchi 753-8512, Japan}

\author[0000-0002-9321-7406]{M.~Hayashi}
\affiliation{National Astronomical Observatory of Japan, 2-21-1 Osawa, Mitaka, Tokyo 181-8588, Japan}

\author[0000-0003-1747-2891]{Y.~Matsuda}
\affiliation{National Astronomical Observatory of Japan, 2-21-1 Osawa, Mitaka, Tokyo 181-8588, Japan}
\affiliation{Graduate University for Advanced Studies (SOKENDAI), Osawa 2-21-1, Mitaka, Tokyo 181-8588, Japan}

\author[0000-0002-3789-770X]{K.~Motogi}
\affiliation{Graduate School of Science and Engineering, Yamaguchi University, Yoshida 1677-1, Yamaguchi, Yamaguchi 753-8512, Japan}

\begin{abstract}

We report the time variability of the late-time radio emission in a Type-I superluminous supernova (SLSN), PTF10hgi, at $z = 0.0987$. 
The Karl G. Jansky Very Large Array 3-GHz observations at 8.6 and 10 years after the explosion both detected radio emission with a $\sim$40\% decrease in flux density in the second epoch. 
This is the first report of a significant variability of the late-time radio light curve in a SLSN. 
Through combination with previous measurements in two other epochs, we constrained both the rise and decay phases of the radio light curve over three years, peaking at approximately 8--9 years after the explosion with a peak luminosity of $L_{\rm 3GHz} = 2 \times 10^{21}$ W~Hz$^{-1}$. 
Possible scenarios for the origin of the variability are an active galactic nucleus (AGN) in the host galaxy, an afterglow caused by the interaction between an off-axis jet and circumstellar medium, and a wind nebula powered by a newly-born magnetar. 
Comparisons with models show that the radio light curve can be reproduced by both the afterglow model and magnetar wind nebula model. 
Considering the flat radio spectrum at 1--15 GHz and an upper limit at 0.6 GHz obtained in previous studies, plausible scenarios are a low-luminosity flat-spectrum AGN or a magnetar wind nebula with a shallow injection spectral index. 

\end{abstract}

\keywords{
\href{http://astrothesaurus.org/uat/2008}{Radio transient sources (2008)}; 
\href{http://astrothesaurus.org/uat/508}{Extragalactic radio sources (508)}; 
\href{http://astrothesaurus.org/uat/1668}{Supernovae (1668)}; 
\href{http://astrothesaurus.org/uat/1340}{Radio continuum emission (1340)}; 
\href{http://astrothesaurus.org/uat/1766}{Very Large Array (1766)}; 
\href{http://astrothesaurus.org/uat/1338}{Radio Astronomy (1338)}
}

\section{Introduction} \label{sec:intro}

Superluminous supernovae (SLSNe) are very bright explosions that are $\sim$10--100 times brighter than ordinary Type Ia and core-collapse supernovae (SNe) \citep[see][for a review]{gal-12}. 
SLSNe are classified into two subclasses according to their spectra: hydrogen-poor SLSNe-I and hydrogen-rich SLSNe-II. 
Due to their huge luminosity and scarcity, the physical nature of SLSNe is still a matter of debate, and SLSNe-I are particularly among the least understood SN populations.
There are many models proposed for progenitors and powering sources for SLSNe-I, such as 
pair-instability SNe \citep[e.g.,][]{woos07}, 
spin-down of a newborn strongly magnetized neutron star \citep[magnetar; e.g.,][]{kase10}, 
fallback accretion onto a compact remnant \citep{dext13}, 
and interaction a with dense circumstellar medium (CSM) \citep[e.g.,][]{chev11}.

Late-time radio observations are useful to constrain the models of SLSNe. 
It is expected that radio emission arise from shock interaction between SN ejecta CSM. 
The connection between SLSNe-I and long-duration gamma-ray bursts (GRBs) has been suggested observationally and theoretically \citep[e.g.,][]{lunn14, grei15a, metz15}, and it is thought that an afterglow from an off-axis jet is observable in late-time at radio bands. 
Non-detections in previous studies constrained physical properties, such as energies, mass-loss rates, and CSM densities, for off-axis jets \citep[e.g.,][]{nich16a, marg18b, copp18}. 
Based on the the model of a SN driven by a young pulsar or magnetar \citep{mura16}, \cite{oman18} predicted quasi-state synchrotron radio emission peaking at $\gtrsim$10 years after SN explosion, which can be tested with current radio telescopes. 
Radio observations by \cite{hats18} put constraint on the predictions for one of the SLSNe by \cite{oman18}.

Recently, \cite{efte19} found an unresolved radio source coincident with the position of the SLSN-I SN~2010md/PTF10hgi. 
PTF10hgi was discovered on May 15, 2010 \citep{quim10} in a dwarf galaxy at $z = 0.0987$ \citep{inse13, lunn14, lelo15}. 
Spectral energy distribution (SED) analysis by \cite{perl16} and \cite{schu18} obtained a star-formation rate (SFR) of $\sim$0.1--0.2 $M_{\odot}$~yr$^{-1}$ and stellar mass of $\log{M_*} = 7.6$--7.9 $M_{\odot}$. 
\cite{efte19} argued that the radio emission is consistent with an off-axis jet or wind nebula powered by a magnetar, suggesting the presence of a central engine. 
Further detections at 1.2, 3, and 15 GHz by \citet{law19} and \citet{mond20} support the model of a magnetar-powered SLSN.

In this Letter, we report the time variability in the late-time radio emission of PTF10hgi based on our new 3-GHz radio continuum observations using the Karl G. Jansky Very Large Array (VLA). 
The remainder of the Letter is organized as follows. 
Section~\ref{sec:observations} describes the radio observations, and the results are presented in Section~\ref{sec:results}. 
In Section~\ref{sec:discussion}, we discuss the possible scenarios for the origin of the radio emission. 
Conclusions are presented in Section~\ref{sec:conclusions}. 
Throughout the Letter, we adopt a \cite{chab03} initial mass function and cosmological parameters based on the {\sl Planck} 2018 results \citep{plan20f}. 
The luminosity distance to PTF10hgi is 469 Mpc, and $1''$ corresponds to 2.07 kpc.

\section{VLA Observations} \label{sec:observations}

The VLA S-band 3-GHz (13-cm) observations were performed on December 2, 2018 in semester 18B (Project ID: 18B-077) and April 25, 2020 in semester 20A (Project ID: 20A-133) as part of a search for late-time radio emissions from SLSNe (B. Hatsukade et al., in preparation). 
The observation dates are 8.6 and 10 years after the discovery date of PTF10hgi, respectively. 
The observations were conducted in array configuration C with the baseline length ranging from 45 m to 3.4 km.
The WIDAR correlator was used with 8-bit samplers. 
We used two basebands, each with a 1-GHz bandwidth, centered at 2.5 GHz and 3.5 GHz, which provided a total bandwidth of 2 GHz. 
The SN positions were used as phase centers of the observations.
Bandpass and amplitude calibrations were conducted with 3C286, and phase calibrations were conducted with J1640+1220. 
The details are summarized in Table~\ref{tab:observations}.

The data were reduced with Common Astronomy Software Applications \citep[CASA;][]{mcmu07} release 5.6.2. 
The maps were produced with the task {\tt\string tclean}. 
Briggs weighting with {\tt\string robust} 0.5 was adopted. 
The absolute flux accuracy was estimated by comparing the measured flux density of the amplitude calibrator and the flux density scale of \cite{perl17}, and the difference was found to be $<$1\%.

\begin{deluxetable*}{lcccccccccc}
\tablecaption{VLA 3-GHz Observations and Results \label{tab:observations}}
\tablewidth{0pt}
\tabletypesize{\footnotesize}
\tablehead{
\colhead{Date} & \colhead{Semester} & \colhead{Time since explosion} 
& \colhead{$T_{\rm on}$$^a$} & \colhead{$N_{\rm ant}$$^b$} 
& \colhead{Beam size$^c$} & \colhead{P.A.$^c$} & \colhead{RMS$^d$} & \colhead{$S_{\rm 3 GHz}$$^e$} & \colhead{$L_{\rm 3 GHz}$} \\
\colhead{} & \colhead{} & \colhead{(yr)} & \colhead{(min)} & \colhead{} 
& \colhead{(arcsec)} & \colhead{($^{\circ}$)} & \colhead{($\mu$Jy beam$^{-1}$)} & \colhead{($\mu$Jy)} & \colhead{(W Hz$^{-1}$)}
}
\startdata
2018-12-02 & 18B &  8.6 &  91 & 26 & $ 8.9 \times 5.8$ & 42.7 & 5.3 & $85 \pm 7$ & $(2.0 \pm 0.2) \times 10^{21}$ \\
2020-04-25 & 20A & 10.0 & 103 & 28 & $ 7.8 \times 6.3$ & 43.5 & 5.8 & $51 \pm 6$ & $(1.2 \pm 0.1) \times 10^{21}$ \\
\enddata
\tablecomments{
$^a$ On-source integration time. 
$^b$ Number of antennas. 
$^c$ Synthesized beam size and position angle. 
$^d$ RMS noise level of the map.  
$^e$ The emission was spatially unresolved and we adopt the peak intensity. 
The uncertainty is the combination of the map rms and a 5\% absolute flux calibration uncertainty
\footnote{https://science.nrao.edu/facilities/vla/docs/manuals/oss/performance/fdscale}.}
\end{deluxetable*}

\begin{figure}
\begin{center}
\includegraphics[width=.9\linewidth]{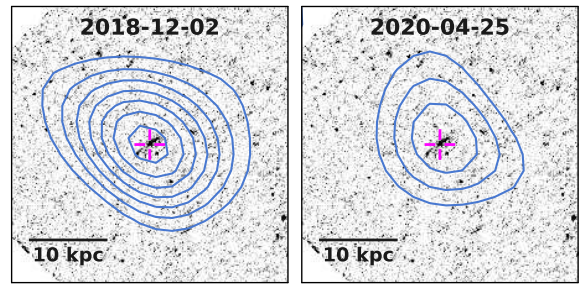}
\end{center}
\caption{
VLA 3-GHz contours of PTF10hgi obtained on November 2, 2018 (left) and April 25, 2020 (right) overlaid on the host galaxy image taken with {\sl HST} WFC3/UVIS F336W. 
North is up, and east is to the left. 
The contours start from $3\sigma$ with $2\sigma$ increments. 
The crosses represent the SN position. 
The scale bar shows 10 kpc (4\farcs9) at the distance of PTF10hgi. 
\label{fig:map}}
\end{figure}

\section{Results} \label{sec:results}

\begin{figure*}
\begin{center}
\plottwo{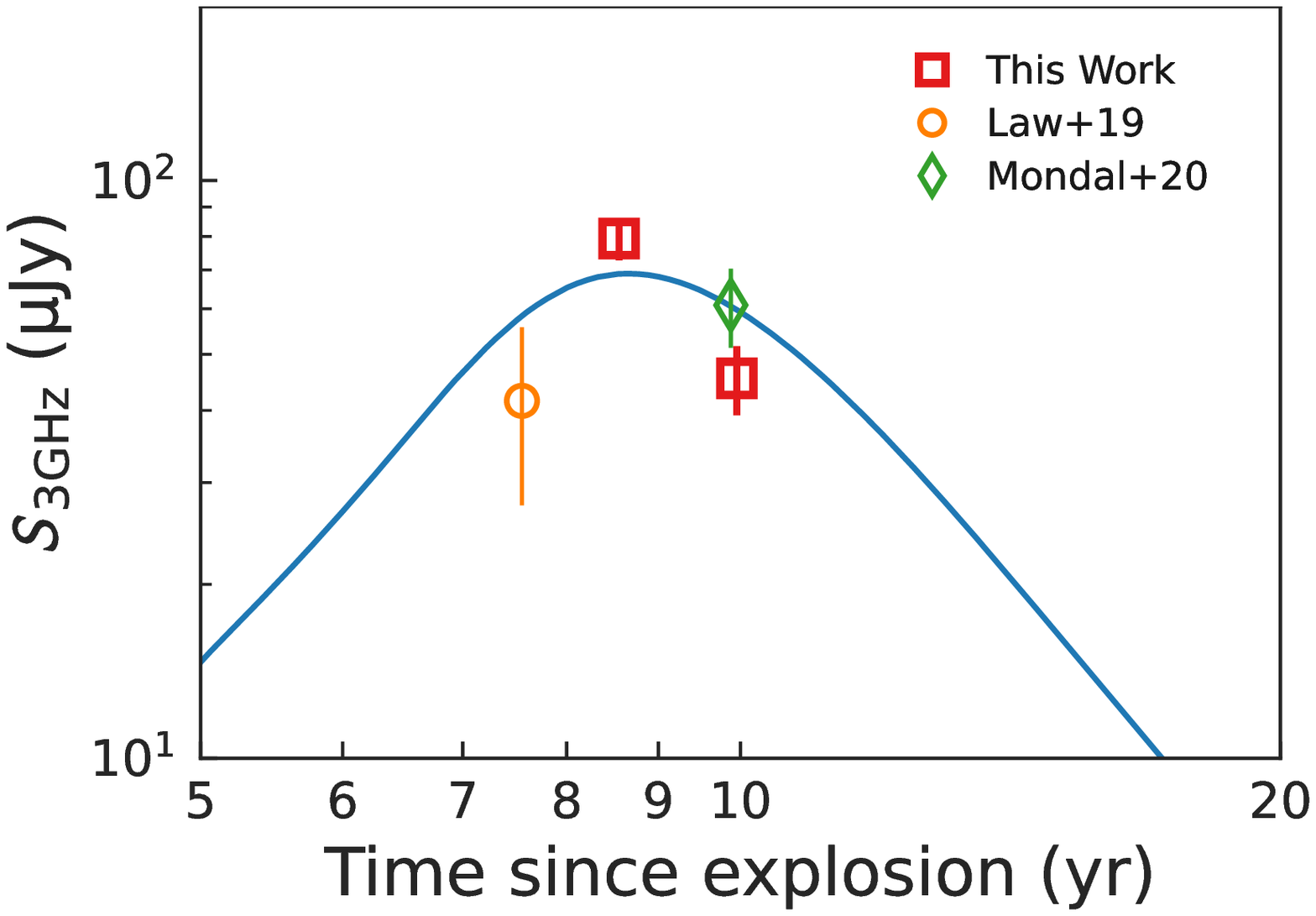}{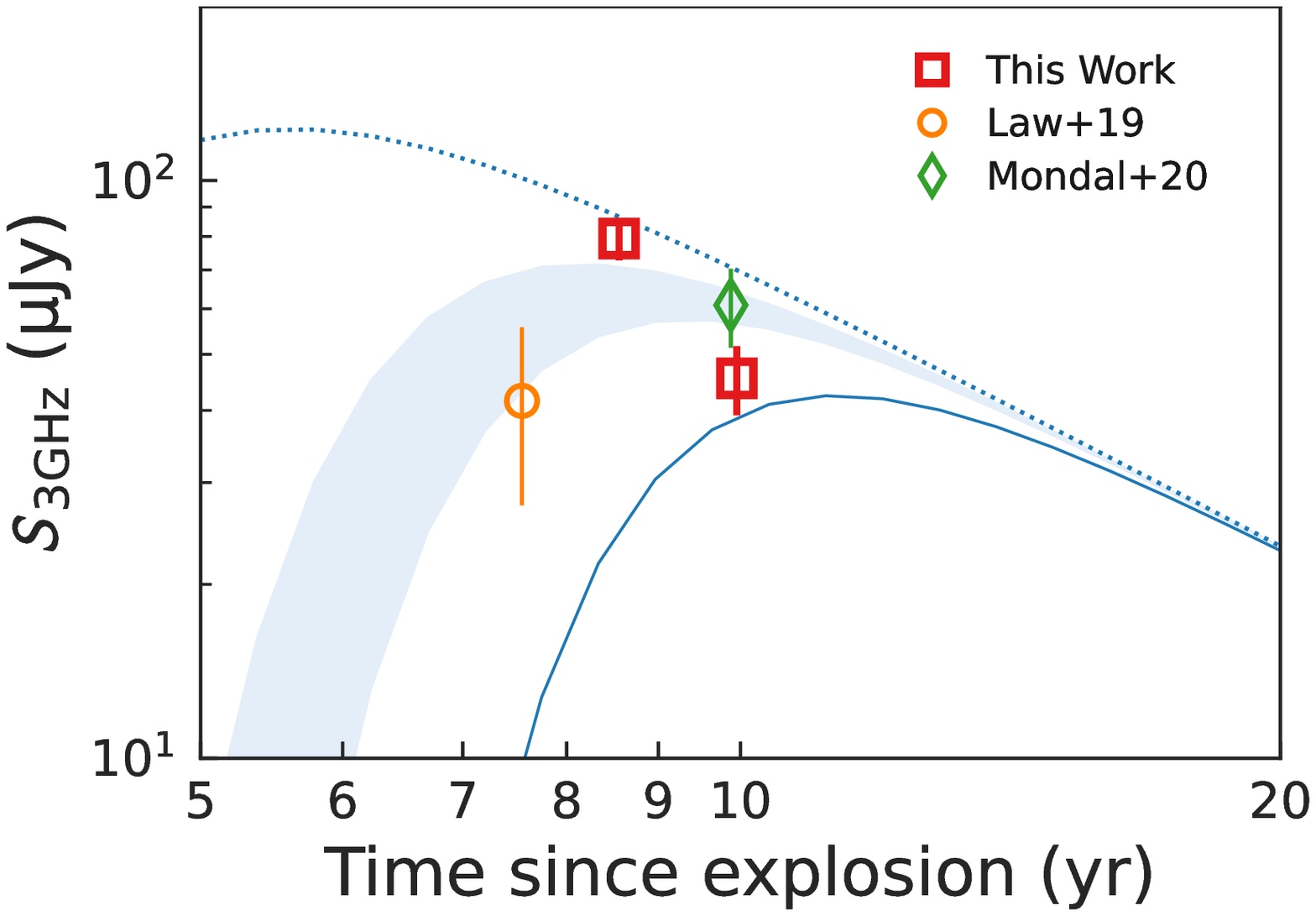}
\end{center}
\caption{
Light curve of PTF10hgi at 3 GHz. 
Our observed data are presented as squares along with the results of \cite{law19} at 3 GHz and \cite{mond20} at 3.3 GHz. 
The radio emission expected from the star-forming activity in the host galaxy is subtracted from the data points (see text). 
Left panel: 
We plot an afterglow model generated using the {\tt afterglowpy} code \citep{ryan20} with $E_{\rm iso} = 1.5 \times 10^{54}$ erg and $n = 7 \times 10^{-3}$ cm$^{-3}$. 
Right panel: 
We plot the magnetar model presented by \cite{law19} with and without free-free absorption as solid and dotted lines, respectively, scaled by a factor of 1.6. 
The shaded region shows the model with 40--60\% of the ejecta singly ionized. 
}
\label{fig:lightcurve}
\end{figure*}

The 3-GHz images taken in the two semesters are shown in Figure~\ref{fig:map}. 
Radio emission was significantly detected in both semesters with peak signal-to-noise ratios of 16 and 9 in the former and latter semesters, respectively. 
The positional uncertainties of peak position were estimated to be $\sim$0\farcs3 and $\sim$0\farcs5, respectively. 
The peak position is consistent with the SN location and with previous observations at 3 and 6 GHz within the positional uncertainties \citep{efte19, law19}. 
We conducted a 2-D Gaussian fit to the emission in the image plane, which could not deconvolve source from the synthesized beam in both semesters. 
Because the emission was spatially unresolved in the observations, we adopted the peak intensity as a source flux density. 
We found a significant time variability of the radio emission between the two semesters with a 40\% decrease in flux density. 
Note that we found no systematic difference in flux densities of $>$$5\sigma$ sources detected in the same field-of-view in the two semesters. 
In Figure~\ref{fig:lightcurve}, we plot the flux densities as a function of time together with two previous measurements at 3 GHz by \cite{law19} and \cite{mond20}. 
The light curve shows both the rise and decay phases over three years with a peak luminosity of $(2.0 \pm 0.2) \times 10^{21}$ W~Hz$^{-1}$. 
We searched for other radio observations of PTF10hgi in wide-field radio surveys and in the literature. 
\cite{schu18} reported a non-detection in the 1.4 GHz data of the NRAO VLA Sky Survey \citep[NVSS;][]{cond98} with a nominal rms level of $\sim$0.45 mJy~beam$^{-1}$. 
We also did not find radio emission in the 3 GHz image of the Very Large Array Sky Survey \citep[VLASS;][]{lacy20} observed in March 2019 with an rms noise level of $\sim$0.22 mJy~beam$^{-1}$. 
These noise levels are relatively shallow compared to the observations by \cite{efte19}, \cite{law19}, \cite{mond20}, and in this study, and do not provide useful constraints on the light curve. 
By using the four data points with radio detection at 3 GHz, we calculated the flux coefficient of variation \cite[$V$;][]{swin15a}, which is equivalent to the fractional variability, defined as $V = s / \bar{F}$, where $s$ is the standard deviation of the flux measurements and $\bar{F}$ is the mean flux density. 
The calculated fractional variability was $V = 0.24$, corresponding to 24\% variability in flux density. 
The maximum-to-median flux density ratio was 1.5. 
This is the first case of a significant time variability being reported in the late-time radio light curve of a SLSN.

\section{Discussion} \label{sec:discussion}

There are various possibilities for the origin of the variability of radio emission. 
It is possible that the effects of scintillation could have resulted in the flux changes \citep[e.g.,][]{rick90}. 
If it were to be the case and the cause of the higher flux density for the first semester, the intrinsic luminosity of the source would be nearly constant. 
However, it is hard to quantify the effects of scintillation with only four data points. 
In what follows, we assume the variability is intrinsic to the source and discuss the physical origin related to the SLSN or its host galaxy.

\cite{efte19} discussed possible origins for the radio emission, such as star formation activity in the host galaxy, an active galactic nucleus (AGN), interaction between the SN ejecta/jet and CSM, and magnetar wind nebulae. 
The significant time variability we found in this study enables us to reject the steady radio emission from star-formation activity. 
We discuss the possibilities of an AGN, shock interaction, and magnetar wind nebulae in the following sections.

\subsection{AGN} \label{sec:agn}

The peak position of the radio emission coincides with the SN location and with the center of the host galaxy (Figure~\ref{fig:map}). 
Although the optical line diagnostic based on a BPT \citep{bald81} diagram shows that the host galaxy lies on the star-forming branch \citep{lelo15, perl16}, it has been reported that the diagram is biased against AGNs in low-mass, blue star-forming galaxies \citep[e.g.,][]{trum15}. 
Deep radio surveys have revealed extragalactic variable sources and AGN signatures in faint radio sources \citep[e.g.,][]{mool16, radc19, smol17, alge20a, rein20}. 
\cite{sarb20a} conducted a deep blind survey of radio variables at 1-2 GHz probing down to faint sources ($<$100 $\mu$Jy) on timescales ranging from days to years. 
They found that 4.9\% of radio sources (18 out of 370) have fractional variabilities of $V > 0.1$, and their host galaxies show AGN signatures based on thresholds such as X-ray luminosity, mid-infrared colors, optical-to-mm SEDs, and radio excess. 
Deep radio continuum surveys found that faint sources with radio luminosities or stellar masses similar to those of the PTF10hgi host show AGN features based on various criteria \citep{smol17, alge20a}. 
A sensitive search for radio emission towards 111 dwarf galaxies ($M_* < 3 \times 10^9$ $M_{\odot}$) by \cite{rein20} found that 13 galaxies have compact radio sources that are almost certainly AGNs.

\begin{figure*}
\begin{center}
\plottwo{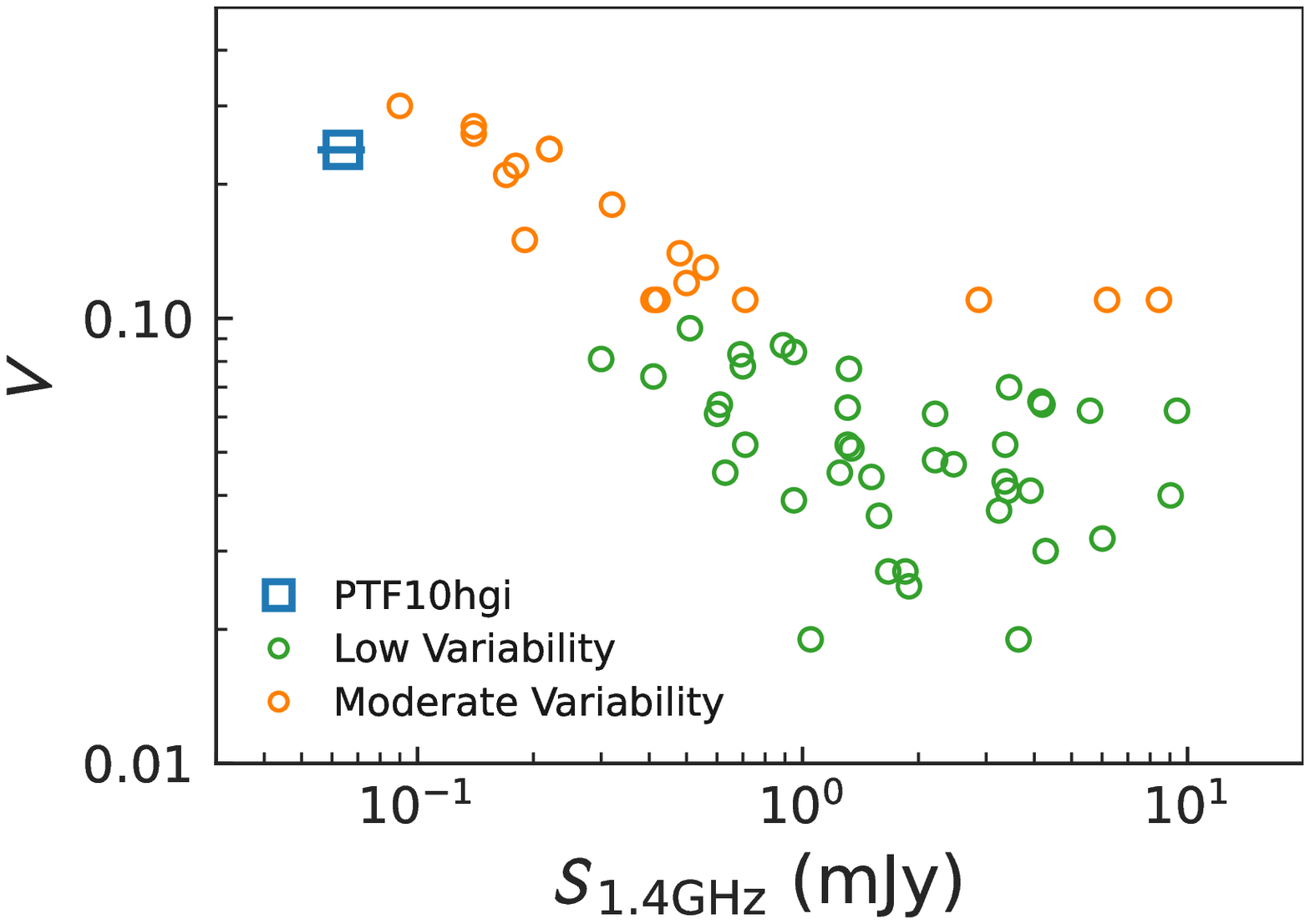}{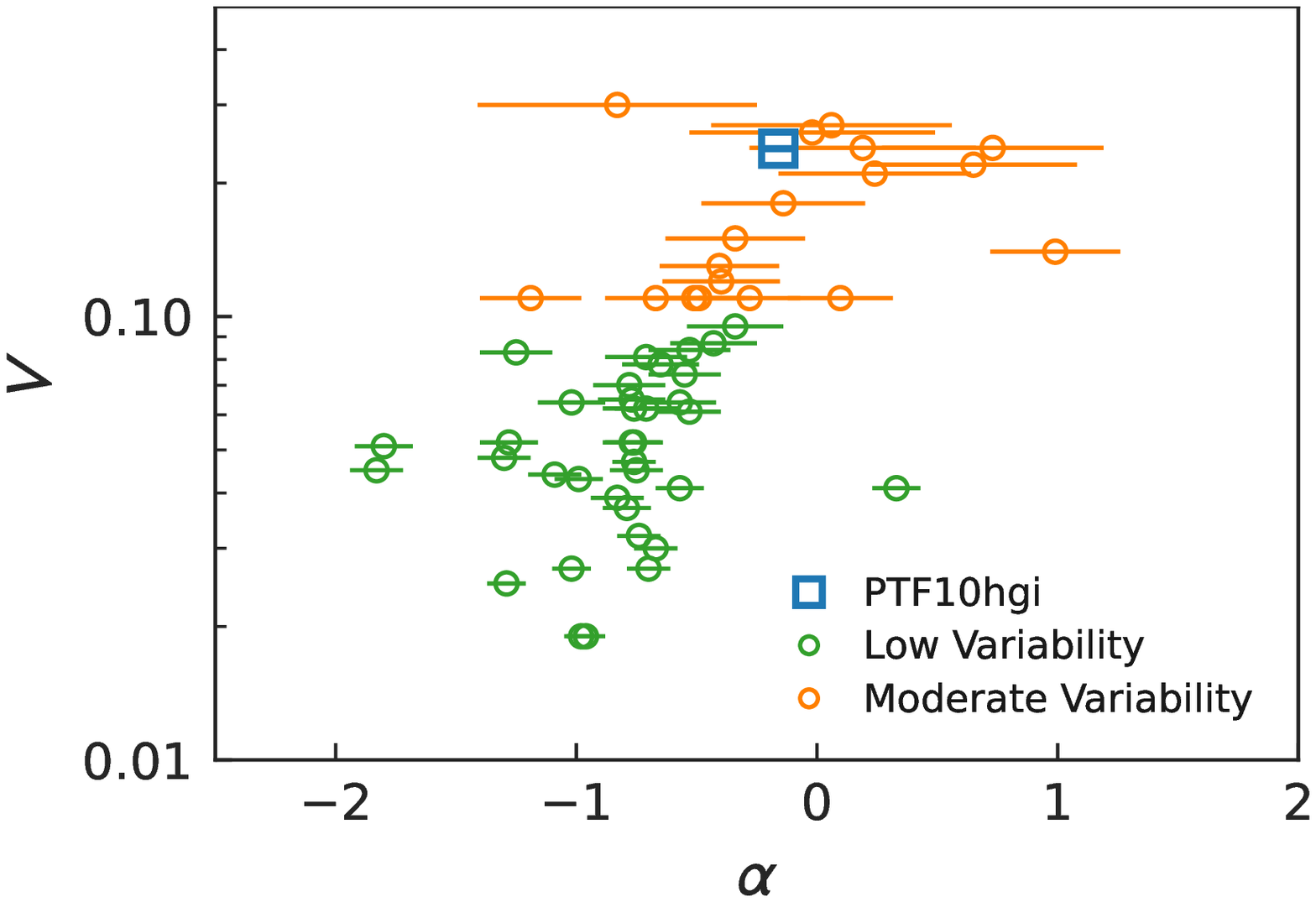}
\end{center}
\caption{
Left panel: 
Comparison between 1.4 GHz flux density and flux coefficient of variation $V$. 
The flux density of PTF10hgi is taken from the measured by \citet{mond20} at 1.2 GHz. 
Radio variable sources with `moderate-variability' ($V > 0.1$) and `low-variability' ($V < 0.1$) found by \citet{sarb20a} are plotted for comparison.  
Right panel: 
Comparison between radio spectral index $\alpha$ and $V$. 
The spectral index was measured at 1.2--15 GHz for PTF10hgi and at 1.4--3 GHz for the sample of radio variable sources by \citet{sarb20a}. 
}
\label{fig:variability}
\end{figure*}

Figure~\ref{fig:variability} compares $V$ with flux density or radio spectral index $\alpha$ (defined as $S_{\nu} \propto \nu^{\alpha}$). 
The radio spectral index of PTF10hgi was calculated to be $\alpha = -0.14 \pm 0.06$ using the data points between 1.2 and 15 GHz obtained in this study and by \citet{efte19}, \citet{law19}, and \citet{mond20}. 
Note that we subtracted the expected radio contribution due to star-forming activity from the data points when driving the spectral index, because the radio observations did not spatially resolve the host galaxy. 
We calculated the radio emission expected from the SFR of 0.15 $M_{\odot}$~yr$^{-1}$, which is in between the estimates of \cite{perl16} and \cite{schu18}: 9.2, 5.4, 3.8, and 2.5 $\mu$Jy at 1.2, 3, 6, and 15 GHz, respectively. 
Figure~\ref{fig:variability} shows that PTF10hgi shares a similar region to those of moderate-variability sources ($V > 0.1$) found in the survey of \citet{sarb20a}. 
\citet{mond20} obtained a flat spectrum of PTF10hgi in the frequency range 1.2--15 GHz and an upper limit at 0.6 GHz. 
Figure~\ref{fig:spectrum} shows the radio spectrum of PTF10hgi. 
The spectral features are similar to those of flat-spectrum sources or gigahertz-peaked-spectrum (GPS) sources, which have radio spectra with peak frequencies of a few GHz \citep[][for a review]{ode98}. 
Their spectral shape may arise from compact cores and jets due to self-absorbed synchrotron emission. 
It was found that the linear source size is negatively correlated with turnover frequency in their spectra \citep{ode98}, giving a linear size of $\sim$100-500 pc for sources with peaks at $\sim$1--3 GHz, which is consistent with the upper limit obtained in 6-GHz observations of PTF10hgi \citep[$\lesssim$2 kpc;][]{efte19}. 
These sources are thought to represent the early stages of the evolution of radio AGN, confinement to small spatial scales by a dense interstellar medium, or intermittent phases of AGN activity. 
They are typically bright radio sources, but it is expected that a large population of low-luminosity sources exist, which are yet to be explored \citep[e.g.,][]{coll18}.

\subsection{Interaction between SN Outflow and CSM} \label{sec:sn}

\cite{efte19} discussed the external shock interaction between SN outflow and CSM as the origin of radio emission. 
They considered two scenarios: non-relativistic (quasi-)spherical SN ejecta and an off-axis relativistic jet. 
For the spherical SN ejecta scenario, they investigated the ejecta velocity and mass-loss rate based on the phase-space of peak luminosity versus peak time assuming that the radio observations were conducted around the peak time. 
The inferred values are significantly different from those of known radio-emitting SNe Ib/c, and they concluded that this scenario is unlikely.

The other scenario is that the radio emission is an afterglow arising from an initially off-axis jet that decelerated and spread into the line of sight at late times. 
They generated afterglow models for a range of jet energies and CSM densities and found that the observed 6-GHz flux density measured approximately 7.5 years after the explosion can be reproduced with an isotropic-equivalent energy $E_{\rm iso} \sim$ (3-5) $\times 10^{53}$ erg and CSM densities $n \sim 10^{-3}$--$10^2$ cm$^{-3}$. 
Now that we have more data points that show the time variability, we can examine whether the light curve can be explained by afterglow models. 
We utilize the publicly available {\tt afterglowpy} code \citep{ryan20}, which generates afterglow light curves using semianalytic approximations of the jet evolution and synchrotron emission. 
We adopted a top hat jet with following fixed parameters: 
jet opening angle of $10^{\circ}$, 
electron energy distribution index of $p = 2.5$, 
thermal energy fraction in electrons of $\epsilon_e = 0.1$, 
and thermal energy fraction in magnetic field of $\epsilon_B = 0.01$, which are typical values for GRBs \citep[e.g.,][]{wang15}; these values were also assumed by \cite{efte19} and \cite{efte20a}. 
The left panel of Figure~\ref{fig:lightcurve} shows a model light curve along with the radio data. 
We found that the data points can be reproduced with $E_{\rm iso} \approx 10^{54}$ erg, $n \approx 10^{-2}$ cm$^{-3}$, and a viewing angle of $\theta_{\rm obs} = 60^{\circ}$. 
The inferred energy is in the highest range of GRBs \citep[e.g.,][]{butl10, wang15}. 
\cite{copp18} compiled all radio observations of SLSNe-I and constrained energies and mass-loss rates or CSM densities for off-axis jets, but lower-density environments ($n \lesssim 10^{-2}$ cm$^{-3}$) with larger viewing angles were not ruled out. 
However, the spectral index of the afterglow model, $(1 - p)/2$ ($p > 2$), is inconsistent with the observed flat radio spectrum. 
Even if we consider the time evolution of the spectral index of an afterglow \citep{sari98}, it is difficult to reproduce the flat spectrum.

\subsection{Magnetar Wind Nebula} \label{sec:magnetar}

PTF10hgi has been proposed to be a magnetar-powered SLSN \citep{efte19, law19, mond20}. 
\cite{efte19} argued that the properties of the source are consistent with a magnetar wind nebula, and its timescale or luminosity can be reproduced by scaling the magnetar model for the persistent radio source associated with the repeating FRB~121102 \citep{metz17}. 
\cite{law19} detected radio emission in VLA 3-GHz observations and found that the emission is consistent with the interpretation that it is powered by a magnetar with free-free absorption in partially ionized ejecta. 
They calculated the time evolution of radio emission from the pulsar wind nebulae (PWNe) based on the model of \cite{mura15} and \cite{mura16}. 
\cite{mura16} showed that under their model, pulsar-driven SN remnants cause quasi-steady synchrotron radio emission associated with non-thermal electron-positron pairs in nascent PWNe on a timescale of decades. 
Based on this model, \cite{law19} estimated the initial parameters of a magnetar (spin period $P_i$, magnetic field $B$, and ejecta mass $M_{\rm ej}$) by fitting the early optical light curve. 
They assumed an electron-positron injection spectrum motivated by Galactic PWNe, such as the Crab PWN \citep[e.g.][]{tana10, tana13}, a broken power law with a peak Lorentz factor of $\gamma_b = 10^5$, and injection spectral indices of $q_1 = 1.5$ and $q_2 = 2.5$. 
They found that a model with ($P_i$, $B$, $M_{\rm ej}$) = (1 ms, $1.4 \times 10^{13}$ G, 15 $M_{\odot}$) and 30--50\% of the ejecta singly ionized can reproduce the observed data at 3 and 6 GHz.

\begin{figure}
\begin{center}
\includegraphics[width=.9\linewidth]{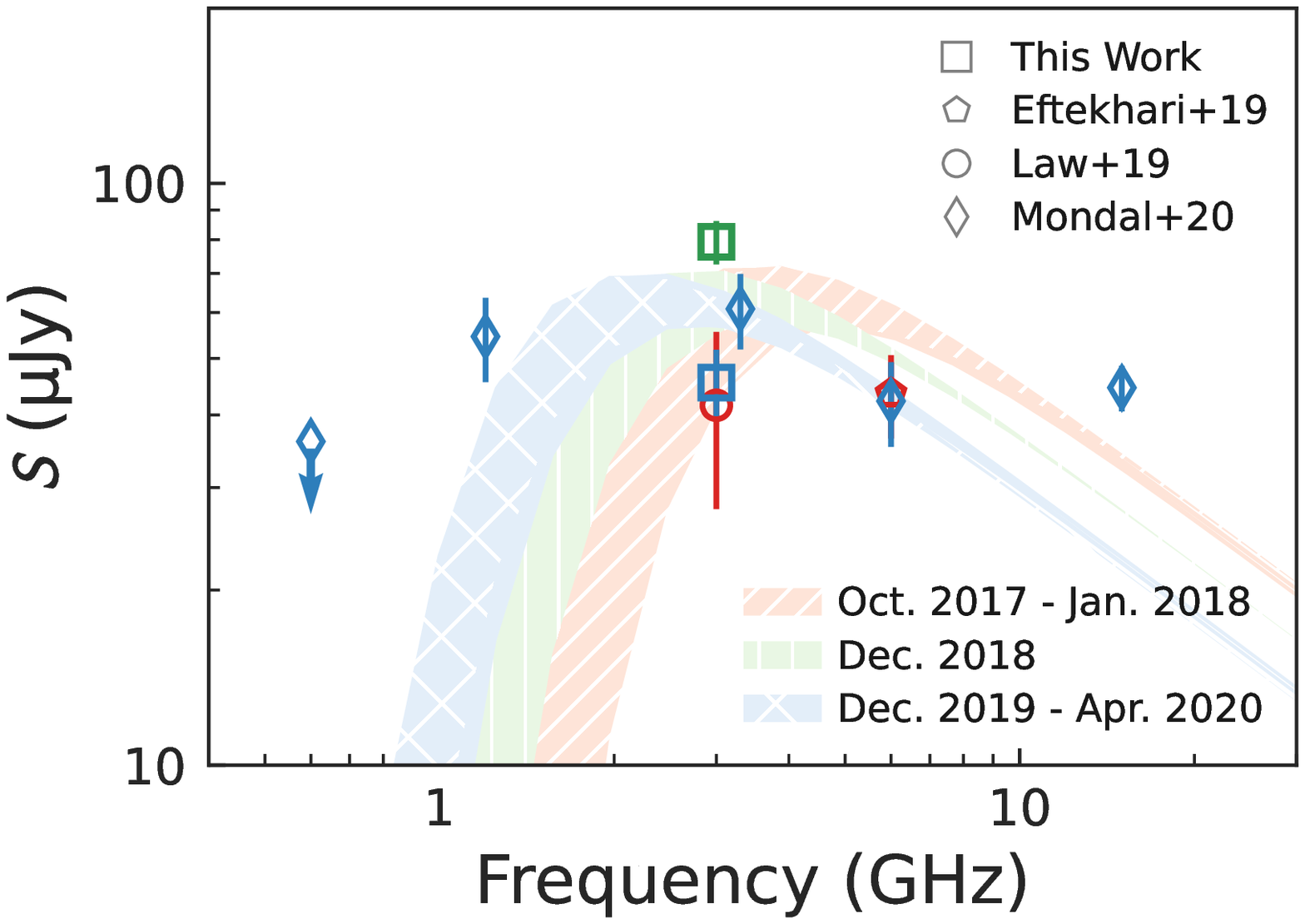}
\end{center}
\caption{
Radio spectra of PTF10hgi. 
The observed data are taken from our results and the literature \citep{efte19, law19, mond20}. 
Radio emission expected from the star-forming activity in the host galaxy was subtracted from the data. 
We plot the magnetar model presented by \cite{law19} scaled by a factor of 1.6 with free-free absorption 40--60\% of the ejecta singly ionized. 
The data points and models are color-coded by their observation epochs. 
}
\label{fig:spectrum}
\end{figure}

In the right panel of Figure~\ref{fig:lightcurve}, we plot the magnetar model presented by \cite{law19} with the same initial parameters. 
We found that this model with 40--60\% of the ejecta singly ionized and scaled by a factor of 1.6 in the vertical direction can reproduce the observed light curve. 
The increase of flux density by a factor of 1.6 can be achieved by a slight modification of the parameters, such as a $\sim$30\% decrease of the magnetic field. 
The timescale of the declining phase appears to be shorter than that of the model. 
One possibility is a rapid spin-down of a young pulsar by the loss of rotational energy due to not only magnetic dipole radiation but also dissipation processes \citep[e.g.,][]{maed07}. 
Alternatively, a steeper spectral injection index would cause a rapid decline. 
However, a larger $q_1$ yields a steeper spectrum, which is inconsistent with the flat spectrum in the range of 1--15 GHz \citep{mond20}. 
Figure~\ref{fig:spectrum} shows the radio spectrum of PTF10hgi along with the same magnetar model as in Figure~\ref{fig:lightcurve} by changing observation epochs. 
We divided the data set into three epochs to see the time variability of the spectral index in Figure~\ref{fig:spectrum}. 
We found no significant change between the epochs of October 2017--January 2018 ($\alpha = 0.07 \pm 0.54$) and December 2019--April 2020 ($\alpha = -0.08 \pm 0.07$), although the power-law fitting results are not stringent. 
It is difficult to compare with the data of December 2018 due to its limited number of points. 
The model significantly underpredicts the 15-GHz flux, as noted by \cite{mond20} and \cite{efte20a}. 
A shallower spectral injection index could be a solution for reproducing the flatter spectrum \citep{mond20}. 
This may suggest a need for modifying the standard model of PWNe \citep[e.g.,][]{ishi17}. 
To constrain the models and parameters, multi-frequency long-term monitoring is required.

\section{Conclusions} \label{sec:conclusions}

We conducted VLA 3-GHz observations of SLSN-I PTF10hgi ($z = 0.0987$) 8.6 and 10 years after its explosion. 
Radio emission was significantly detected in both epochs. 
We found a time variability with a $\sim$40\% decrease in flux density in the second epoch. 
Through combination with previous measurements in two other epochs, we constrained both the rise and decay phases of a radio light curve over three years peaking at approximately 8--9 years after the explosion. 
This is the first report of variability of a late-time radio light curve in a SLSN. 
A possible scenario for the origin of the variability is a low-luminosity AGN in the host galaxy. 
Another possibility is an afterglow caused by the interaction between an off-axis jet and CSM. 
Comparison with models shows that although the light curve can be reproduced, the predicted radio spectrum is inconsistent with the observed flat spectrum. 
Alternatively, we found that the light curve can be reproduced by a magnetar wind nebula model. 
Our findings provide important implications for the central engine of SLSNe.

Current data sets are not enough to make definitive settlements on PTF10hgi, and more data at different times and frequencies are required. 
A decrease in flux density on a long timescale would be favored for the central engine scenarios, because fluctuations of observed flux density can be caused by AGN activity or scintillation. 
While a flat spectrum can be explained by AGN or magnetar scenarios, a steeper spectral index is expected for afterglow models. 
Because a time evolution of spectral index could also happen in magnetar wind nebula models \citep{mura16, oman18}, long-term monitoring with multi-frequencies are important.

\acknowledgments

We would like to acknowledge NRAO staffs for their help in preparation of observations. 
We thank the referee for helpful comments and suggestions which significantly improved the Letter. 
We are grateful to the PDJ collaboration for providing opportunities for fruitful discussions. 
BH is supported by JSPS KAKENHI Grant Number 19K03925. 
The National Radio Astronomy Observatory is a facility of the National Science Foundation operated under cooperative agreement by Associated Universities, Inc.
Based on observations made with the NASA/ESA Hubble Space Telescope, obtained from the Data Archive at the Space Telescope Science Institute, which is operated by the Association of Universities for Research in Astronomy, Inc., under NASA contract NAS 5-26555. 
This research has made use of the CIRADA cutout service at URL cutouts.cirada.ca, operated by the Canadian Initiative for Radio Astronomy Data Analysis (CIRADA). CIRADA is funded by a grant from the Canada Foundation for Innovation 2017 Innovation Fund (Project 35999), as well as by the Provinces of Ontario, British Columbia, Alberta, Manitoba and Quebec, in collaboration with the National Research Council of Canada, the US National Radio Astronomy Observatory and Australia's Commonwealth Scientific and Industrial Research Organisation.

\facility{VLA}


\listofchanges
\end{document}